\documentclass[useAMS,usenatbib]{mn2e}

\usepackage{amssymb,amsmath}
\usepackage{graphicx}
\usepackage{url}
\usepackage{ulem}
\usepackage{longtable}
\usepackage{multicol}

\newcommand{\be}{\begin{equation}}
\newcommand{\ee}{\end{equation}}
\newcommand{\ben}{\begin{equation*}}
\newcommand{\een}{\end{equation*}}

\usepackage{color}

\title[Direct photodissociation]
{Non-thermal photons and direct photodissociation of H$_2$, HD and HeH$^{+}$ in the chemistry of the primordial Universe}

\author[C.~M.~Coppola, M.~V.~Kazandjian, D.~Galli, A.~N.~Heays, E.~F.~van~Dishoeck]
{C.~M. Coppola$^{1,2}$
\thanks{e-mail: carla.coppola@uniba.it}, M.~V.~Kazandjian$^{3}$, D.~Galli$^{2}$, A.~N.~Heays$^{3,4}$, E.~F.~van~Dishoeck$^{3,5}$ \\
$^1$Universit\`{a} degli Studi di Bari, Dipartimento di Chimica, Via Orabona 4, I-70126, Bari, Italy\\
$^2$INAF-Osservatorio Astrofisico di Arcetri, Largo E.~Fermi 5, I-50125 Firenze, Italy\\
$^3$Leiden Observatory, Leiden University, PO Box 9513, 2300 RA, Leiden, The Netherlands\\
$^4$LERMA, Observatoire de Paris, PSL Research University, CNRS, Sorbonne Universit\'es, UPMC Univ. Paris 06, F-92190, Meudon, France\\
$^5$Max-Planck-Institut fur extraterrestriche Physik, Postfach 1312, 85741, Garching, Germany\\}

\begin{document}

\date{}

\maketitle

\begin{abstract}
Non-thermal photons deriving from radiative transitions among the internal ladder of atoms and molecules are an important source of photons in addition to thermal and stellar sources in many astrophysical environments. In the present work the calculation of reaction rates for the direct photodissociation of some molecules relevant in early Universe chemistry is presented; in particular, the calculations include non-thermal photons deriving from the recombination of primordial hydrogen and helium atoms for the cases of H$_2$, HD and HeH$^+$. New effects on the fractional abundances of chemical species are investigated and the fits for the HeH$^+$ photodissociation rates by thermal photons are provided. 
\end{abstract}


\section{Introduction}
Photodissociation processes represent important channels to destroy molecules in several astrophysical
environments; the mechanisms through which they occur and effect on the chemistry deeply depend on
the chemical species involved and on the features of the radiation field these molecules are embedded in. For example, photodissociation 
due to UV photons produced by the interaction of cosmic-rays with dense interstellar clouds
has been reported in the literature \citep{prasad1983,sternberg1987, heays2017} and the effects of such UV flux on the 
chemistry has been described for several chemical species \citep{gredel1987,gredel1989,heays2014}.
Another example is represented by the X-ray spectra emitted by high-mass young stellar objects (YSOs) that
are usually fitted with the emission spectrum of an optically thin thermal plasma
\citep{hofner1997} and that has been used to describe the chemistry in the envelopes around YSOs \citep{stauber2006}.
On the other hand, photodissociation can occur following specific dynamical 
pathways according to the features of the potential energy surfaces describing the possible electronic states
of the molecules themselves. In particular, direct photodissociation,
predissociation and spontaneous radiative photodissociation are described as the main
ways to photodestroy small molecules 
(e.g. contribution by van Dishoeck in \citealt{vandishoeck1988} and \citealt{vandishoeck_book_2015});
according to the dynamics, different features in the photodissociation 
cross sections can be observed.
Eventually, considering both the radiation field and the dynamics of the photodissociation processes,
additional terms can be calculated other than the thermal emission contribution to the reaction rate of
photo-processes.

When moving to the early Universe case, non-thermal radiation fields can arise as a result of different
mechanisms such as matter/antimatter annihilation, decaying relic/dark matter particles,
dissipation of acoustic waves \citep[see e.g.][]{chluba2011a}, radiative cascade of H$_2$ \citep{coppola2012}
among others. They are also called ``distortion photons''
because of their departure from the Planckian shape of the Cosmic Microwave Background spectrum (hereafter CMB). Among these additional radiation
fields, the most relevant for its effect on the chemistry of the primordial Universe and for the level of accuracy
with which it has been modelled is represented by the spectrum deriving from the recombination processes
of H and He in the so-called epoch of recombination (EoR) \citep{chluba2010,chluba2011b}. In fact, because of the adiabatic expansion of the Universe and the effective Compton scattering, the temperature of the matter dropped, allowing for the first bound atomic states to form. The effect of non-thermal photons on the early Universe chemistry has been studied in several papers \citep{hirata2006}. In particular \cite{coppola2013} 
presented a modified version of the chemistry in the primordial Universe, where the effect of
non-thermal photons was investigated on two photodestruction processes that are relevant for the
chemistry of H$_2$, namely the photodissociation of H$_2^+$ and the photo-detachment of H$^-$.

In the present work we focus on the effect of non-thermal photons
on the photodissociation of the molecular species that are 
of interest for the primordial Universe. The calculations presented 
by \cite{coppola2013} will be here extended to the direct photodissociation
of H$_2$, HD and HeH$^+$.
Although quite simple systems, they represent the key molecular species present in the early Universe chemistry;
indeed while H$_2$ and HD are connected to the cooling of the gas down to few tens of kelvins in the
low-metallicity environment present at high redshifts \citep[e.g., ][]{galli1998,lepp2002,dalgarno2005},
HeH$^+$ contributes to the opacity and optical properties of the primordial gas itself
\citep[e.g., ][]{schleicher2008}.

This paper is organized as follows: in Section~\ref{methods} the formalism used for the description and
implementation of non-thermal photons in the chemical kinetics is introduced, and the processes investigated are listed. The quantum dynamical features of each channel are described, and references for the
used cross sections are provided. In Section~\ref{results} the resulting non-thermal rate coefficients 
are shown and the effects on the chemical kinetics are discussed and reported.

\section{Formulation of the problem}
\label{methods}

\subsection{Distortion photons and non-thermal rate coefficient}
\label{subsection1}
Radiative transitions in any quantum system between higher, $i$, and lower, $j$, internal energy levels (or a lower-energy continuum) are
associated with the emission of a photon, causing a spectral distortion of specific intensity $\Delta  I_{ij}(\nu)$.
The observed frequency, $\nu$, of a photon emitted at redshift $z_{\rm em}$ and observed at redshift $z$ is related to its rest frame frequency, $\nu_{ij}$ according to $\nu=\nu_{ij}(1+z)/(1+z_{\rm em})$, assuming a narrow line profile. The spectral distortion produced by the emission process at $z_{\rm em}$ and observed at redshift $z < z_{\rm em}$ can be written \citep[e.g., ][]{rubino2008}:
\be
\Delta I_{ij}^z(\nu)=\left(\frac{hc}{4\pi}\right)
\frac{\Delta R_{ij}(z_{\rm em})(1+z)^3}{H(z_{\rm em})(1+z_{\rm em})^3}
\label{e2}
\ee
where $H(z)=H_0[\Omega_{\rm r}(1+z)^4+\Omega_{\rm m}(1+z)^3+
\Omega_{\rm k}(1+z)^2+\Omega_\Lambda]^{1/2}$
is the Hubble function and $\Delta R_{ij}$ is related to the
population of the $i^{\rm th}$ and $j^{\rm th}$ levels by:
\be
\Delta R_{ij} = p_{ij} A_{ij} N_i 
\frac{e^{h\nu_{ij}/k_{\rm B}T_{\rm r}}}{e^{h\nu_{ij}/k_{\rm B}T_{\rm r}}-1}
\left[1-\frac{g_i N_j}{g_j N_i}e^{-h\nu_{ij}/k_{\rm B}T_{\rm r}}\right],
\ee
where $p_{ij}$ is the Sobolev-escape probability, $g_i$ and $g_j$ 
the degeneracy of upper and lower levels, respectively (both factors are equal
to one in the case of pure vibration transitions), $A_{ij}$ is the Einstein coefficient of the transition, and
$T_{\rm r}=2.726~(1+z_{\rm em})\,$K \citep{fixsen2009}. 

\begin{figure}
\includegraphics[width=8.5cm]{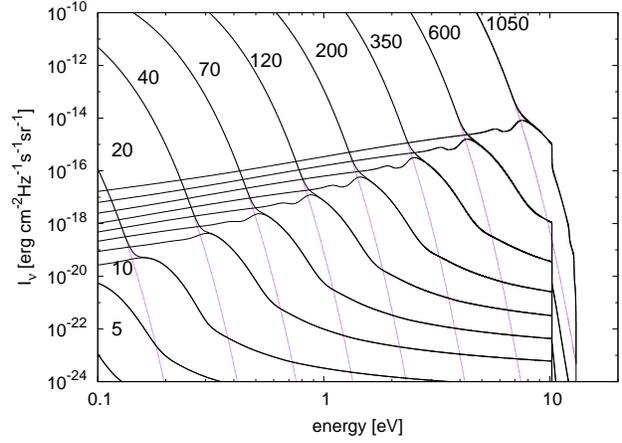}
\caption{Thermal and non-thermal spectra at different redshift $z$. Together with the Planck distribution at different radiation temperatures (purple dotted curves, corresponding to the redshift values reported in the figure), the non-thermal contributions are reported. The latter derive both from primordial atomic recombination of H and He and H$_2$ radiative cascade.}
\label{spectra}
\end{figure}

Eventually, the total contribution of spectral distortions to
the rate of a reaction with photons at a given redshift, $z$, can be evaluated by integration
over the photon distribution:
\be
k_{\rm ph}(z)=4 \pi \int_0^\infty \frac{\sigma(\nu)}{h\nu} 
\left[ B_z(\nu)+\sum_{i\rightarrow j} \Delta I_{ij}^z(\nu)\right] {\rm d} \nu.
\label{rc}
\ee
In Eq.~\ref{rc}, $\sigma(\nu)$ represents the cross section of the process as a
function of frequency, $B_z(\nu)$ is the Planck distribution at
$T_{\rm r}$ corresponding to the redshift $z$ at which the reaction
rate is calculated. In Fig.~\ref{spectra}
the spectra for the non-thermal and thermal photons are reported for different values of the redshift; the calculations performed in this paper have been based on these spectra. Additional details on non-thermal photons distribution can be found in \cite{coppola2013}.

\subsection{Molecular species}

In the following, a description of the photodissociation processes and molecular species modeled in the present work is given.

\subsubsection{H$_2$}
The photodissociation of H$_2$ proceeds by two 
dynamical mechanisms; first, the Solomon process consists of a two-step pathway with bound-bound
resonant absorption through the Lyman and Werner bands followed by fluorescent
decay into the continuum of the ground electronic state:
\begin{equation}
\mathrm{H}_2^{(X; v,j)} + h\nu \rightarrow \mathrm{H}_2^{(B/C;v',j')} \rightarrow \mathrm{H(1s)} + \mathrm{H(1s)} + h\nu
\label{solomon}
\end{equation}
The cross sections show a peculiar peaked behaviour at the energy corresponding to the energies of the emitted photons \citep[e.g.,][]{mentall1977}.
The Solomon process has been always treated as the main photodestruction channel 
of H$_2$ in studies of several environments \citep{stecher1967,abgrall1992,abgrall2000}, including the early Universe case
\citep{galli1998,lepp2002,dalgarno2005}. Although reaction \ref{solomon} represent the main channel
though which photodissociation occurs it has been shown that the direct continuum photodissociation process
\begin{equation}
\mathrm{H}_2^{(X; v,j)} + h\nu \rightarrow \mathrm{H(1s)} + \mathrm{H(2p)}
\label{directh2}
\end{equation}
can effect the total rate coefficients, for example, in the case of interstellar clouds \citep{shull1978}
and in the early Universe chemistry \citep{coppola2011}.
Several authors have calculated the cross sections for process \ref{directh2} 
\citep{allison1969,glass1986,zucker1986}; more recently \cite{gay2012} provided rovibrationally 
resolved cross sections. They are available at the website http://www.physast.uga.edu/ugamop/,  together with the energy levels. The overall H$_2$ photoexcitation-emission cross section is highly structured, being a mixture of Doppler-limited line emission and an underlying continuum due to processes (4) and (5), respectively.

\subsubsection{HD}
HD photodissociation is homologous to the H$_2$ case. A cross section is calculated by  \cite{allison1969},
though only vibrationally resolved. In the present work, the rotational quantum number of the available cross sections is assumed to be equal to zero.

\subsubsection{HeH$^+$}
The photodissociation of HeH$^+$ has been extensively studied both from a theoretical and experimentalist point
of view (\citealt{loreau2011,urbain2012,gay2012}), also in the case of excited electronic states \citep{loreau2013, miyake2011}. In this case, the photodissociation is dominated by two processes:
\begin{equation}
\mathrm{HeH}^+ (X ^1\Sigma^+, v, J) + h\nu\rightarrow \mathrm{He}^+(1s)\mathrm{+ H}(1s) (A ^1\Sigma^+)
\end{equation}

\begin{equation}
\mathrm{HeH}^+ (X ^1\Sigma+, v, J) + h\nu \rightarrow \mathrm{He}(1s^2) + \mathrm{H}^+ (X ^1\Sigma^+)
\end{equation}

In the following, these are referred to as A~$\leftarrow$~X and X~$\leftarrow$~X photodissociation, respectively. As in the case of direct photodissociation of H$_2$, rovibrationally resolved cross sections and energy levels are available at the website http://www.physast.uga.edu/ugamop/; the details on the calculations can be found in the work by \cite{miyake2011}. These two channels have been previously inserted in chemical networks describing the formation and destruction of primordial molecules \citep[e.g.,][]{galli1998, lepp2002, schleicher2008}.

\section{Results}
\label{results}
The main results concern both photodissociation rates and fractional abundances in the context of early Universe chemistry. In the following the results are described according to the chemical species.
\subsection{Reaction rates}
\subsubsection{H$_2$ and HD}
\begin{figure}
\begin{center}$
\begin{array}{l}
\includegraphics[width=8.5cm]{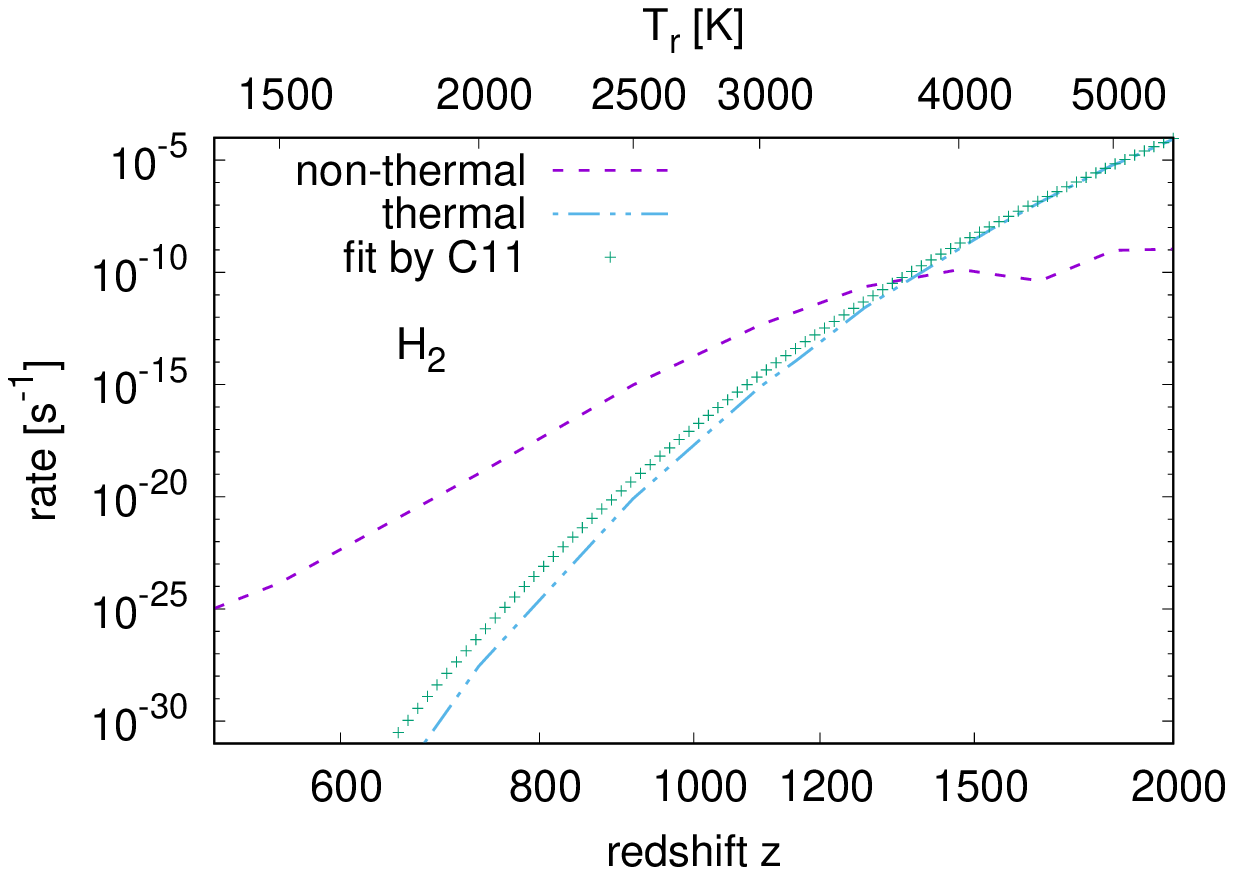}\\
\includegraphics[width=8.5cm]{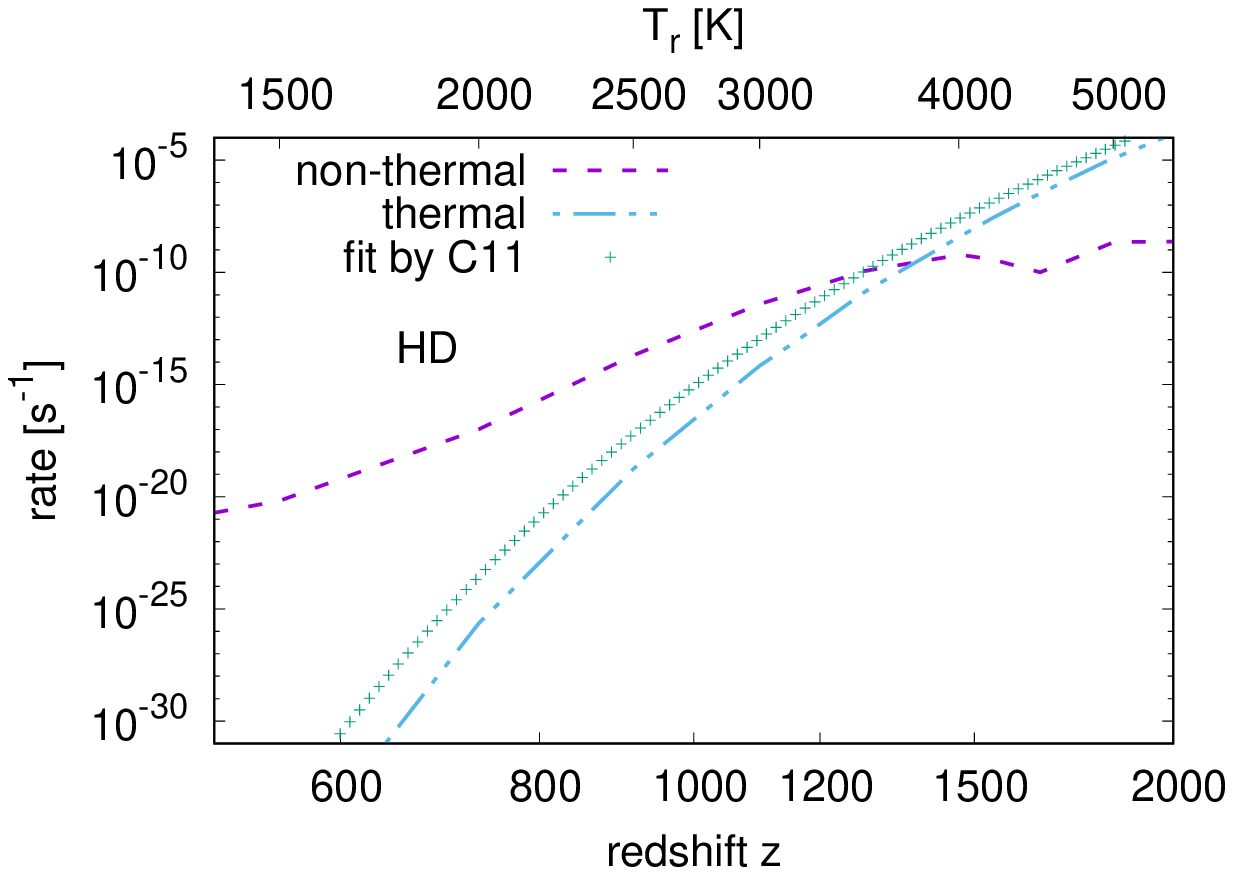}\\
\end{array}$
\end{center}
\caption{Photodissociation rates as a function of redshift for direct photodissociation of H$_2$ (top panel) and HD (bottom panel). Both rates calculated adopting a thermal radiation spectrum and non-thermal photons are shown. Comparison with data calculated by Coppola et al.~(2011a) (reported as C11 in the key) is provided.} 
\label{h2_hd_reactionrate}
\end{figure}

In Fig.~\ref{h2_hd_reactionrate}, thermal and non-thermal reaction rates for the process of direct photodissociation calculated according to Eq.~\ref{rc} are shown, for H$_2$ and HD (top and bottom panels, respectively) as a function of redshift. The presence of extra-photons produced by the primordial recombination of H and He results in the formation of a non-thermal tail in the reaction rate. The value of the redshift at which the crossing-over between non-thermal and the thermal reaction rate appears is $z \sim 1300$; at this $z$, the radiation temperature is T$_r \sim 3545~K$. The curves referred to as C11 correspond to the fits provided by \cite{coppola2011}.

\subsubsection{HeH$^+$}
\begin{figure}
\begin{center}$
\begin{array}{l}
\includegraphics[width=8.5cm]{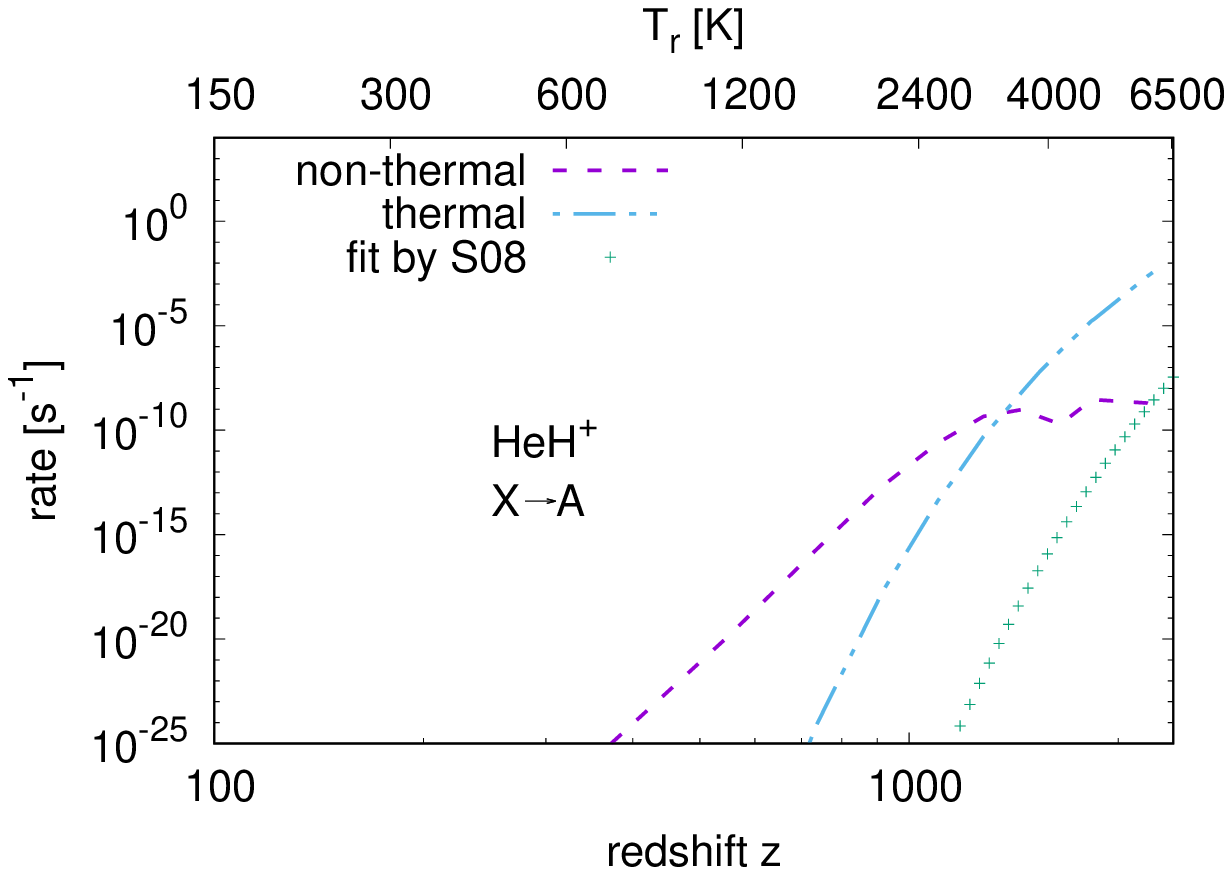}\\
\includegraphics[width=8.5cm]{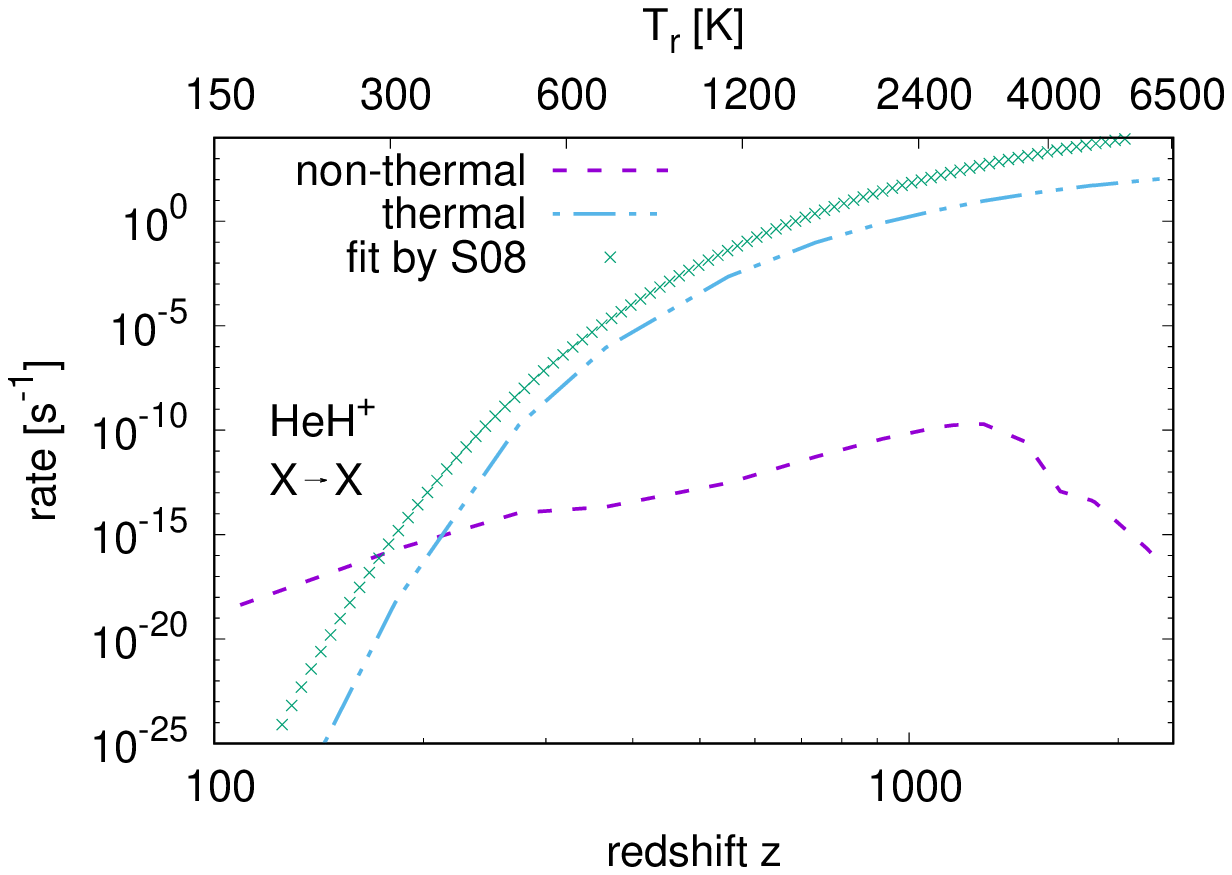}\\
\end{array}$
\end{center}
\caption{Photodissociation rates for HeH$^+$: thermal and non-thermal contribution. {\it Top panel}: 
transition A$\leftarrow$X; {\it bottom panel}: transition X$\leftarrow$X. The blue curve is the 
thermal contribution calculated in the present work adopting the cross sections by Gay et al.~2012 while 
the green curve represents the contribution of non-thermal photons to the reaction rate. 
The crosses represent fits implemented by Schleicher et al.~(2008)} 
\label{hehp_reactionrates}
\end{figure}

In Fig.~\ref{hehp_reactionrates} the reaction rates for direct photodissociation of HeH$^+$ are shown, separately for the cases of A~$\leftarrow$~X and X~$\leftarrow$~X. It is possible to see two important features: firstly, photodissociation rates from thermal photons are quite different from the usually-adopted fits (e.g. \citealt{schleicher2008} (referred to as S08 in the figure) and \citealt{galli1998}), which were derived by  detailed balance on the data for radiative association of He and H$^+$ and H and He$^+$. Secondly, redshift values at which photodissociation rates from non-thermal photons become greater than the thermal contribution are significantly different in the two cases; in the case of the process A~$\leftarrow$~X this departure happens at $z \sim 1100$ and at $z \sim 200$ for the process X~$\leftarrow$~X. Then, the radiation temperatures are quite different, respectively T$_r~\sim~3000~K$ and T$_r~\sim~550~K$.

\subsection{Fractional abundances}
The calculated direct photodissociation rates calculated by considering both thermal and non-thermal emission with the available cross sections have been implemented in a time-dependent chemical network (e.g. \citealt{galli1998}, \citealt{c11}, \citealt{lincei11}, \citealt{galli2013}). The presence of non-thermal photons does not significantly affect the fractional abundances of the chemical species of interest; this result is qualitatively expected from comparing the values at which the departure from thermal to non-thermal features occurs and the maxima in the cross sections.

\begin{figure}
\begin{center}$
\begin{array}{l}
\includegraphics[width=8.5cm]{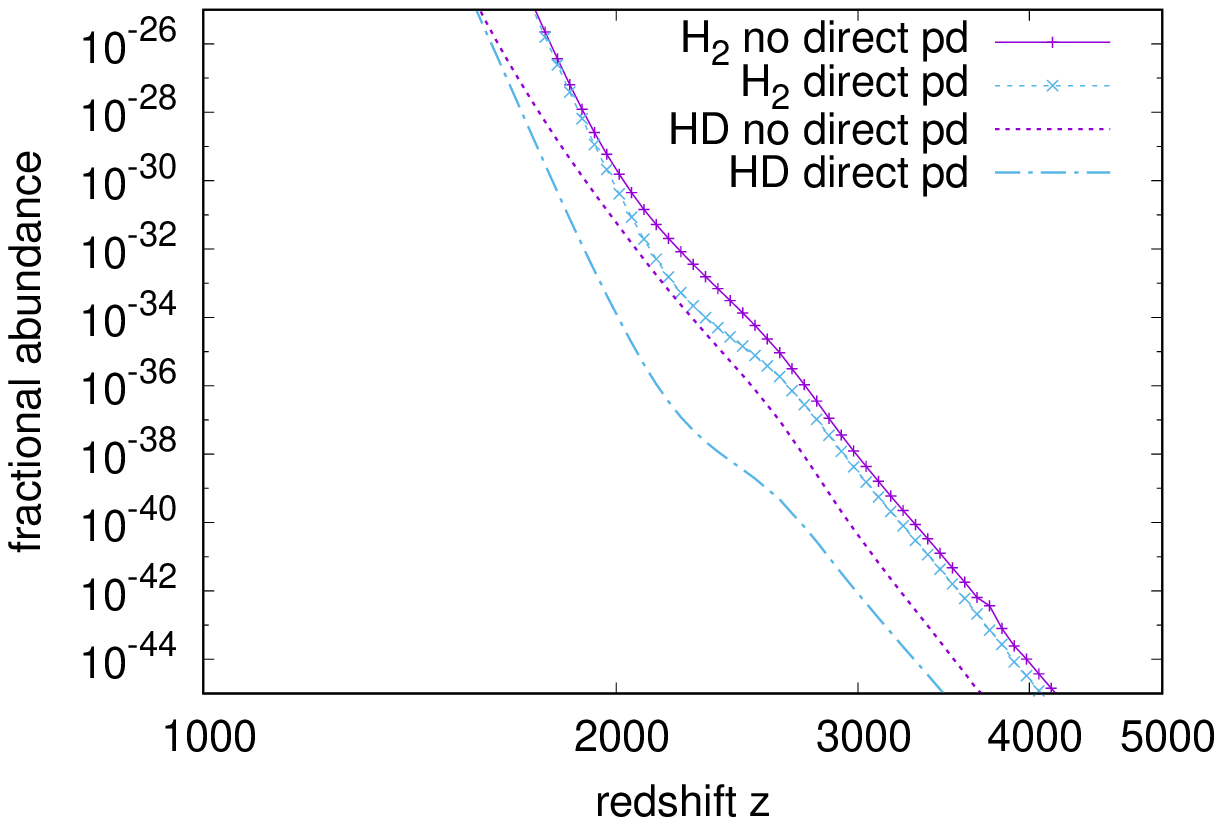}\\
\includegraphics[width=8.5cm]{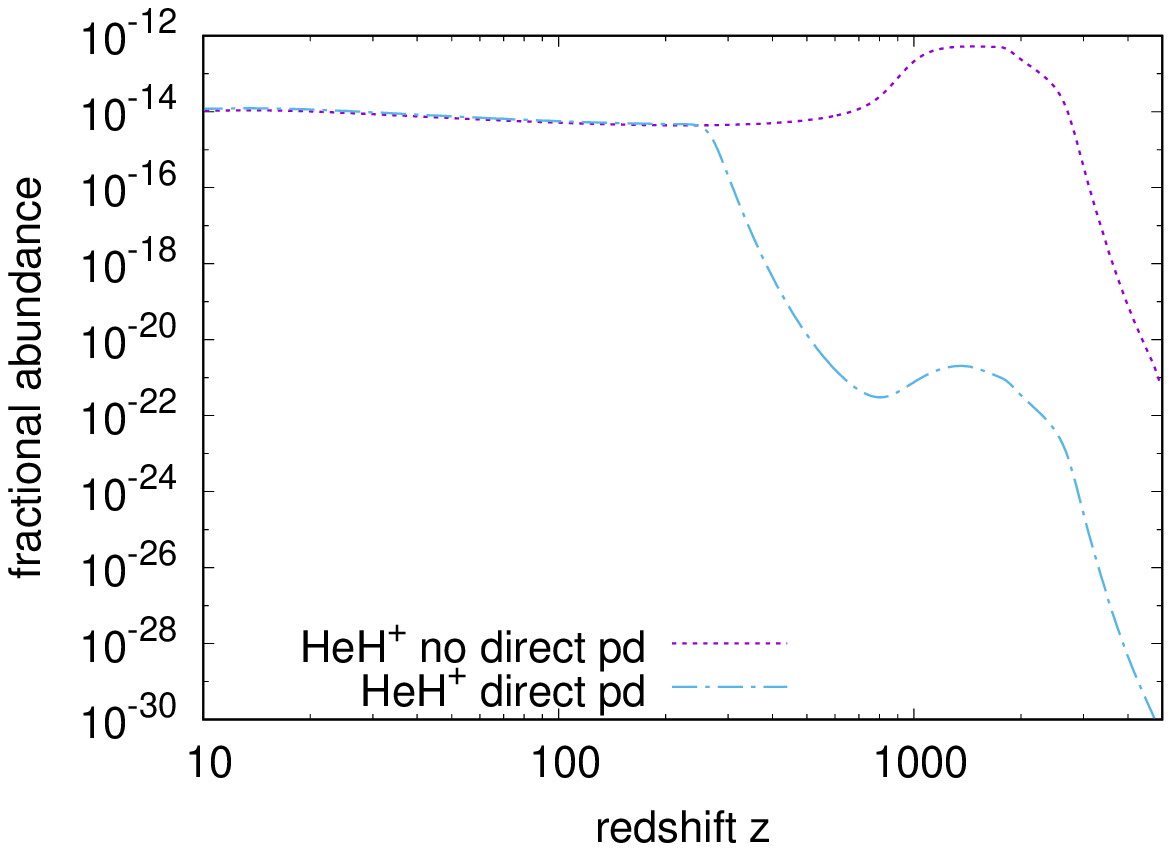}\\
\end{array}$
\end{center}
\caption{Fractional abundances with and without the contribution of direct photodissociation; {\it top panel:}
H$_2$ and HD, {\it bottom panel:} HeH$^+$.} 
\label{fractional_abundances}
\end{figure}

A significant deviation follows from the introduction of the direct process of photodissociation for H$_2$ and HD (in addition to the Solomon processes) at high values of redshift, where differences up to four orders of magnitude can be appreciated (see Fig.~\ref{fractional_abundances}). Although significant, this result does not effect the successive phases of chemical evolution, that are mainly controlled by formation processes occurring at lower redshifts (H$_2^+$ channel followed by the H$^-$ one).

\section{Conclusions}
In the present work the effect of non-thermal photons on the direct photodissociation of three molecules in the context of early Universe chemistry has been investigated. There is no effect on molecular abundances at low redshifts, but some large changes occur at higher $z$. Such effects agree with estimates performed by taking into account on one hand the thresholds for these chemical processes and, on the other hand, the radiation temperature at which they are expected to become significant. In the case of H$_2$ and HD, for example, the energy threshold is quite high; consequently, the direct photodissociation is expected to play a role at high values of $z$, as confirmed by the present simulations. Moreover, the contribution to the photodissociation rates from thermal and non-thermal photons has been provided, showing the ranges at which each term dominates. Updated fits for the direct photodissociation rates of HeH$^+$ as a function of radiation temperature are provide in Appendix A.

\section*{Acknowledgments}
C. M. C. and D. G. acknowledge 
the discussions within the international team \#272 lead by C. M.
Coppola ``EUROPA - Early Universe: Research on Plasma Astrochemistry''
at ISSI (International Space Science Institute) in Bern. C. M. C also greatly acknowledges Regione
 Puglia for the project ``Intervento cofinanziato dal Fondo di Sviluppo e Coesione
 2007-2013 – APQ Ricerca Regione
Puglia - Programma regionale a sostegno della specializzazione intelligente e
della sostenibilit\`a sociale ed ambientale - FutureInResearch''.
\appendix

\section{Fits}
The calculations reported in this paper concerning HeH$^+$ have been performed using the cross sections of \cite{miyake2011}. For convenience, we provide an empirical form of the thermal rates, both for the transition A~$\leftarrow$~X and X~$\leftarrow$~X to the analytical expression:
\begin{equation}
\mathrm{k}(T_r) = a ~ T_r^b ~ \exp{(-c/T_r)}.
\end{equation}
The values of the parameters in both cases are reported in Tab.~\ref{fits}.
\begin{table*}
 \centering
 \begin{minipage}{140mm}
  \caption{HeH$^+$ direct photodissociation reaction rates: updated fits for the thermal contribution as a function
  of T$_r$}
  \begin{tabular}{@{}ll@{}}
  \hline
                                   &               Thermal [s$^{-1}$]    \\
 \hline
$\mathrm{HeH}^+ (X ^1\Sigma^+, v, J) + h\nu\rightarrow \mathrm{He}^+(1s)\mathrm{+ H}(1s) (A ^1\Sigma^+)$    &  \\
                           & a = 273518  \\
                           & b = 0.623525 \\
                           & c = 144044~[K] \\
  \hline
$\mathrm{HeH}^+ (X ^1\Sigma+, v, J) + h\nu \rightarrow \mathrm{He}(1s^2) + \mathrm{H}^+ (X ^1\Sigma^+)$ & \\ 
                           & a = 2.03097 $\times 10^8$  \\
                           & b = -1.20281 \\
                           & c = 24735~[K]   \\
  \hline
\label{fits}
\end{tabular}
\end{minipage}
\end{table*}

\bibliographystyle{mn2e}
\bibliography{biblio}

\begin{thebibliography}{39}
\expandafter\ifx\csname natexlab\endcsname\relax\def\natexlab#1{#1}\fi

\bibitem[{{Abgrall} {et~al}\mbox{.}(1992){Abgrall}, {Le Bourlot}, {Pineau des
  For\^ets}, {Roueff}, {Flower}, \& {Heck}}]{abgrall1992}
{Abgrall} H., {Le Bourlot} J., {Pineau des For\^ets} G., {Roueff} E., {Flower}
  D.~R., {Heck} L., 1992, Astronomy \& Astrophysics, 253, 525

\bibitem[{{Abgrall}, {Roueff} \& {Drira}(2000){Abgrall}, {Roueff}, \&
  {Drira}}]{abgrall2000}
{Abgrall} H., {Roueff} E., {Drira} I., 2000, Astronomy \& Astrophysics
  Supplement, 141, 297

\bibitem[{Allison \& Dalgarno(1969)}]{allison1969}
Allison A.~C., Dalgarno A., 1969, Atomic Data, 1, 91

\bibitem[{{Chluba} \& {Sunyaev}(2012)}]{chluba2011a}
{Chluba} J., {Sunyaev} R.~A., 2012, Monthly Notices of the Royal Astronomical
  Society, 419, 1294

\bibitem[{Chluba \& Thomas(2011)}]{chluba2011b}
Chluba J., Thomas R.~M., 2011, Monthly Notices of the Royal Astronomical
  Society, 412, 748

\bibitem[{Chluba, Vasil \& Dursi(2010)Chluba, Vasil, \& Dursi}]{chluba2010}
Chluba J., Vasil G.~M., Dursi L.~J., 2010, Monthly Notices of the Royal
  Astronomical Society, 407, 599

\bibitem[{Coppola {et~al}\mbox{.}(2012)Coppola, D'Introno, Galli, Tennyson, \&
  Longo}]{coppola2012}
Coppola C.~M., D'Introno R., Galli D., Tennyson J., Longo S., 2012, The
  Astrophysical Journal Supplement Series, 199, 16

\bibitem[{Coppola {et~al}\mbox{.}(2011a)Coppola, Diomede, Longo, \&
  Capitelli}]{coppola2011}
Coppola C.~M., Diomede P., Longo S., Capitelli M., 2011a, The Astrophysical
  Journal, 727, 37

\bibitem[{Coppola {et~al}\mbox{.}(2013)Coppola, Galli, Palla, Longo, \&
  Chluba}]{coppola2013}
Coppola C.~M., Galli D., Palla F., Longo S., Chluba J., 2013, Monthly Notices
  of the Royal Astronomical Society, 434, 114

\bibitem[{{Coppola} {et~al}\mbox{.}(2011b){Coppola}, {Longo}, {Capitelli},
  {Palla}, \& {Galli}}]{c11}
{Coppola} C.~M., {Longo} S., {Capitelli} M., {Palla} F., {Galli} D., 2011b, The
  Astrophysical Journal Supplement Series, 193, 7

\bibitem[{Dalgarno(2005)}]{dalgarno2005}
Dalgarno A., 2005, Journal of Physics: Conference Series, 4, 10

\bibitem[{Fixsen(2009)}]{fixsen2009}
Fixsen D.~J., 2009, The Astrophysical Journal, 707, 916

\bibitem[{{Galli} \& {Palla}(1998)}]{galli1998}
{Galli} D., {Palla} F., 1998, Astronomy and Astrophysics, 335, 403

\bibitem[{{Galli} \& {Palla}(2013)}]{galli2013}
{Galli} D., {Palla} F., 2013, Annual Review of Astronomy and Astrophysics, 51,
  163

\bibitem[{Gay {et~al}\mbox{.}(2012)Gay, Abel, Porter, Stancil, Ferland, Shaw,
  van Hoof, \& Williams}]{gay2012}
Gay C.~D., Abel N.~P., Porter R.~L., Stancil P.~C., Ferland G.~J., Shaw G., van
  Hoof P. A.~M., Williams R. J.~R., 2012, The Astrophysical Journal, 746, 78

\bibitem[{Glass-Maujean(1986)}]{glass1986}
Glass-Maujean M., 1986, Phys. Rev. A, 33, 342

\bibitem[{{Gredel}, {Lepp} \& {Dalgarno}(1987){Gredel}, {Lepp}, \&
  {Dalgarno}}]{gredel1987}
{Gredel} R., {Lepp} S., {Dalgarno} A., 1987, The Astrophysical Journal Letters,
  323, L137

\bibitem[{{Gredel} {et~al}\mbox{.}(1989){Gredel}, {Lepp}, {Dalgarno}, \&
  {Herbst}}]{gredel1989}
{Gredel} R., {Lepp} S., {Dalgarno} A., {Herbst} E., 1989, The Astrophysical
  Journal, 347, 289

\bibitem[{{Heays}, {Bosman} \& {van Dishoeck}(2017){Heays}, {Bosman}, \& {van
  Dishoeck}}]{heays2017}
{Heays} A.~N., {Bosman} A.~D., {van Dishoeck} E.~F., 2017, Astronomy \&
  Astrophysics, DOI 10.1051/0004-6361/201628742

\bibitem[{{Heays} {et~al}\mbox{.}(2014){Heays}, {Visser}, {Gredel}, {Ubachs},
  {Lewis}, {Gibson}, \& {van Dishoeck}}]{heays2014}
{Heays} A.~N., {Visser} R., {Gredel} R., {Ubachs} W., {Lewis} B.~R., {Gibson}
  S.~T., {van Dishoeck} E.~F., 2014, Astronomy \& Astrophysics, 562, A61

\bibitem[{Hirata \& Padmanabhan(2006)}]{hirata2006}
Hirata C.~M., Padmanabhan N., 2006, Monthly Notices of the Royal Astronomical
  Society, 372, 1175

\bibitem[{{Hofner} \& {Churchwell}(1997)}]{hofner1997}
{Hofner} P., {Churchwell} E., 1997, The Astrophysical Journal Letters, 486, L39

\bibitem[{Lepp, Stancil \& Dalgarno(2002)Lepp, Stancil, \& Dalgarno}]{lepp2002}
Lepp S., Stancil P.~C., Dalgarno A., 2002, Journal of Physics B: Atomic,
  Molecular and Optical Physics, 35, R57

\bibitem[{Longo {et~al}\mbox{.}(2011)Longo, Coppola, Galli, Palla, \&
  Capitelli}]{lincei11}
Longo S., Coppola C.~M., Galli D., Palla F., Capitelli M., 2011, Rendiconti
  Lincei, 22, 119

\bibitem[{Loreau {et~al}\mbox{.}(2011)Loreau, Lecointre, Urbain, \&
  Vaeck}]{loreau2011}
Loreau J., Lecointre J., Urbain X., Vaeck N., 2011, Phys. Rev. A, 84, 053412

\bibitem[{{Loreau} {et~al}\mbox{.}(2013){Loreau}, {Vranckx},
  {Desouter-Lecomte}, {Vaeck}, \& {Dalgarno}}]{loreau2013}
{Loreau} J., {Vranckx} S., {Desouter-Lecomte} M., {Vaeck} N., {Dalgarno} A.,
  2013, Journal of Physical Chemistry A, 117, 9486

\bibitem[{Mentall \& Guyon(1977)}]{mentall1977}
Mentall J.~E., Guyon P.~M., 1977, The Journal of Chemical Physics, 67, 3845

\bibitem[{{Millar} \& {Williams}(1988)}]{vandishoeck1988}
{Millar} T.~J., {Williams} D.~A., eds., 1988, Astrophysics and Space Science
  Library, Vol. 146, {Rate coefficients in astrochemistry}

\bibitem[{Miyake, Gay \& Stancil(2011)Miyake, Gay, \& Stancil}]{miyake2011}
Miyake S., Gay C.~D., Stancil P.~C., 2011, The Astrophysical Journal, 735, 21

\bibitem[{{Prasad} \& {Tarafdar}(1983)}]{prasad1983}
{Prasad} S.~S., {Tarafdar} S.~P., 1983, The Astrophysical Journal, 267, 603

\bibitem[{Rubi{\~n}o-Mart{\'{\i}}n, Chluba \&
  Sunyaev(2008)Rubi{\~n}o-Mart{\'{\i}}n, Chluba, \& Sunyaev}]{rubino2008}
Rubi{\~n}o-Mart{\'{\i}}n J.~A., Chluba J., Sunyaev R.~A., 2008, Astronomy and
  Astrophysics, 485, 377

\bibitem[{Schleicher {et~al}\mbox{.}(2008)Schleicher, Galli, Palla, Camenzind,
  Klessen, Bartelmann, \& Glover}]{schleicher2008}
Schleicher D. R.~G., Galli D., Palla F., Camenzind M., Klessen R.~S.,
  Bartelmann M., Glover S. C.~O., 2008, Astronomy and Astrophysics, 490, 521

\bibitem[{{Shull}(1978)}]{shull1978}
{Shull} J.~M., 1978, The Astrophysical Journal, 219, 877

\bibitem[{{St{\"a}uber} {et~al}\mbox{.}(2006){St{\"a}uber}, {J{\o}rgensen},
  {van Dishoeck}, {Doty}, \& {Benz}}]{stauber2006}
{St{\"a}uber} P., {J{\o}rgensen} J.~K., {van Dishoeck} E.~F., {Doty} S.~D.,
  {Benz} A.~O., 2006, Astronomy \& Astrophysics, 453, 555

\bibitem[{{Stecher} \& {Williams}(1967)}]{stecher1967}
{Stecher} T.~P., {Williams} D.~A., 1967, The Astrophysical Journal Letters,
  149, L29

\bibitem[{{Sternberg}, {Dalgarno} \& {Lepp}(1987){Sternberg}, {Dalgarno}, \&
  {Lepp}}]{sternberg1987}
{Sternberg} A., {Dalgarno} A., {Lepp} S., 1987, The Astrophysical Journal, 320,
  676

\bibitem[{Urbain {et~al}\mbox{.}(2012)Urbain, Lecointre, Loreau, \&
  Vaeck}]{urbain2012}
Urbain X., Lecointre J., Loreau J., Vaeck N., 2012, Journal of Physics:
  Conference Series, 388, 022107

\bibitem[{van Dishoeck \& Visser(2014)}]{vandishoeck_book_2015}
van Dishoeck E.~F., Visser R., 2014, Molecular Photodissociation, Wiley-VCH
  Verlag GmbH \& Co. KGaA, pp. 229--254

\bibitem[{Zucker \& Eyler(1986)}]{zucker1986}
Zucker C.~W., Eyler E.~E., 1986, The Journal of Chemical Physics, 85, 7180

\end{thebibliography}

\end{document}